# Integrated information and predictive processing theories of consciousness: An adversarial collaborative review

Andrew W. Corcoran[a,b], Andrew M. Haun[c], Reinder Dorman[d], Giulio Tononi[c], Karl J. Friston[e], Cyriel M. A. Pennartz[d,f], TWCF: INTREPID Consortium

[a] Monash Centre for Consciousness and Contemplative Studies, Monash University, Melbourne, Australia

[b] Department of Nutrition, Exercise and Sports, University of Copenhagen, Copenhagen, Denmark

[c] Department of Psychiatry, University of Wisconsin, Madison, USA

[d] Swammerdam Institute for Life Sciences, Center for Neuroscience, University of Amsterdam, Amsterdam, Netherlands

[e] Queen Square Institute of Neurology, University College London, London, UK

[f] Research Priority Program Brain and Cognition, University of Amsterdam, Amsterdam, Netherlands

**ORCID iDs and e-mail addresses**

AWC: https://orcid.org/0000-0002-0449-4883; anco@nexs.ku.dk
AMH: https://orcid.org/0000-0001-9458-8957; andrew.haun@wisc.edu
RD: https://orcid.org/0000-0001-9596-7051; r.dorman@uva.nl
GT: https://orcid.org/0000-0002-3892-4087; gtononi@psychiatry.wisc.edu
KJF: https://orcid.org/0000-0001-7984-8909; k.friston@ucl.ac.uk
CMAP: https://orcid.org/0000-0001-8328-1175; c.m.a.pennartz@uva.nl

**Corresponding author**

Dr. Andrew W. Corcoran, Institut for Idræt og Ernæring, Københavns Universitet, Nørre Allé 51, 2200 København N, Danmark

# Abstract

As neuroscientific theories of consciousness continue to proliferate, the need to assess their similarities and differences – as well as their predictive and explanatory power – becomes ever more pressing. Recently, a number of structured adversarial collaborations have been devised to test the competing predictions of several candidate theories of consciousness. In this review, we compare and contrast three theories being investigated in one such adversarial collaboration: Integrated Information Theory, Neurorepresentationalism, and Active Inference. We begin by presenting the core claims of each theory, before comparing them in terms of the phenomena they seek to explain, the sorts of explanations they avail, and the methodological strategies they endorse. We then consider some of the inherent challenges of theory-testing, and how adversarial collaboration addresses some of these difficulties. The stage is then set for the empirical work to come: first, we outline the key hypotheses to be tested across a series of multi-site experiments; second, we discuss the kinds of observations that would support or challenge each theory; third, we consider how these theories might assimilate or accommodate such observations. Finally, we show how data harvested across disparate experiments (and their replicates) may be formally integrated to provide a quantitative measure of the evidential support accrued under each theory. Besides orienting the reader to the theoretical foundations of our collaboration, this review aims to provide valuable meta-scientific insights into the mechanics of adversarial collaboration and theory-testing in general – including the way theories may be evaluated in terms of the scientific progress they deliver.

# 1 Accelerating research on consciousness through adversarial collaboration

There is perhaps no other field of neuroscience that generates more widespread interest, disagreement, and controversy than the neuroscience of consciousness. Recent work has highlighted both the proliferation of theories of consciousness and the remarkable diversity of their conceptual foundations and explanatory targets (Kuhn, 2024; Mudrik et al., 2025; Northoff & Lamme, 2020; Seth & Bayne, 2022). Indeed, there are now so many theories of consciousness that an increasing number of researchers are turning their attention to the meta-theoretical question of how they might be compared and evaluated (Chis-Ciure et al., 2024; Del Pin et al., 2021; Doerig et al., 2021; Kirkeby-Hinrup, 2024; Negro et al., 2024; Signorelli et al., 2021), with a view to advancing the field via the unification, integration, or elimination of competing candidates (Evers et al., 2024; Storm et al., 2024; Wiese, 2020). It is in this context that the Templeton World Charity Foundation launched their 'Accelerating Research on Consciousness' (ARC) initiative (https://acceleratingresearch.org), an ambitious attempt to progress consciousness science by pitting some of its leading theories directly against one another in a series of structured adversarial collaborations (Melloni et al., 2021; Reardon, 2019).

Adversarial collaboration is a unique approach to scientific inquiry in which proponents of competing theories work together to identify genuine points of disagreement that can be adjudicated by empirical investigation (Kahneman, 2003; Mellers et al., 2001). One or more experiments are then jointly devised by members of each adversarial party to test the contrasting predictions of their theories, with a commitment to publish their findings irrespective of the outcome. While often challenging to implement (Melloni, 2022; Tetlock & Mitchell, 2009a, 2009b; Vlasceanu et al., 2022), the adversarial model confers several notable benefits, including (but not limited to): (1) complementing open science practices with procedures designed to reduce bias and increase validity; (2) incentivising fair but 'severe' tests (Mayo, 2018; Popper, 2002) of competing hypotheses; (3) ensuring that inconvenient findings cannot be easily dismissed due to differences in methodological predilection (Ceci et al., 2024; Clark et al., 2022; Clark & Tetlock, 2023; Rakow, 2022). Such advantages may be particularly apt for driving progress in a field such as consciousness science (Corcoran et al., 2023), where there is widespread disagreement about the way consciousness should be operationalised and the kinds of data that are relevant for theory development and critique (Francken et al., 2022; Irvine, 2017; Yaron et al., 2022).

The first stage of the adversarial collaborative process calls for participants to work together to form a clear understanding of one another's theoretical commitments and disagreements (Bateman et al., 2005; Cowan et al., 2020; Peters et al., 2025). Indeed, a core tenet of this paradigm is the capacity to characterise the position of one's adversary with accuracy and precision, thereby establishing a firm foundation for the joint development of experiments that are capable of arbitrating substantive theoretical differences (Clark et al., 2022; Corcoran et al., 2023). The present 'adversarial review' documents the outcome of this process for the INTREPID Consortium (https://arc-intrepid.com), an adversarial collaboration funded under the ARC initiative to test competing predictions of the Integrated Information Theory and two predictive processing theories of consciousness – Neurorepresentationalism and Active Inference (Figure 1).

In what follows, we present a succinct overview of the three theories at hand, identify notable points of commonality and disagreement amongst them, and examine how their distinctive theoretical claims are translated into hypotheses amenable to empirical investigation. Crucially, the agenda here is not to engage in parochial disputes concerning the merits and shortcomings of one another's theories, but rather to lay the theoretical foundations for the empirical work to come. As part of this agenda, we will consider some of the inherent difficulties of theory testing in general – and in the arena of consciousness science specifically – where the appropriate explanatory targets for inquiry (and methodological strategies for investigating them) remain controversial. We explicate how the INTREPID Consortium addressed this challenge by formulating several complementary experiments designed to test various kinds of hypotheses. We then anticipate how the potential outcomes of these experiments will support or challenge each of the theories involved, before showing how the data generated by disparate experiments may be assimilated under a Bayesian evidence accumulation scheme. We hope that this review will provide a useful resource – not only for those engaged in debates about theories of consciousness (and indeed other topics; see, e.g., https://gac.ccneuro.org/; https://web.sas.upenn.edu/adcollabproject) – but also for those interested in the broader (meta-theoretical) question of how empirical evidence for competing theoretical claims can be accrued, integrated, and evaluated over time.

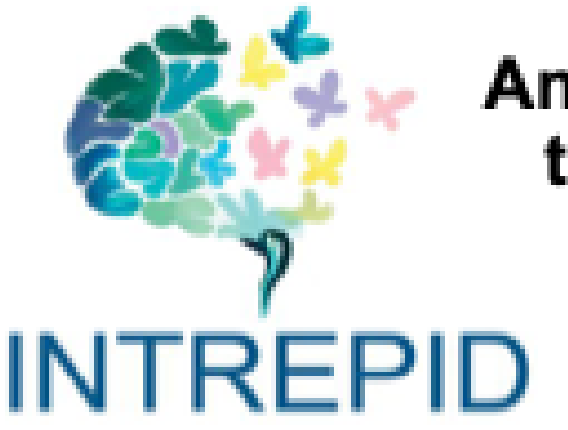

**An adversarial collaboration testing three theories of consciousness**

**Predictive Processing**

## Integrated Information Theory

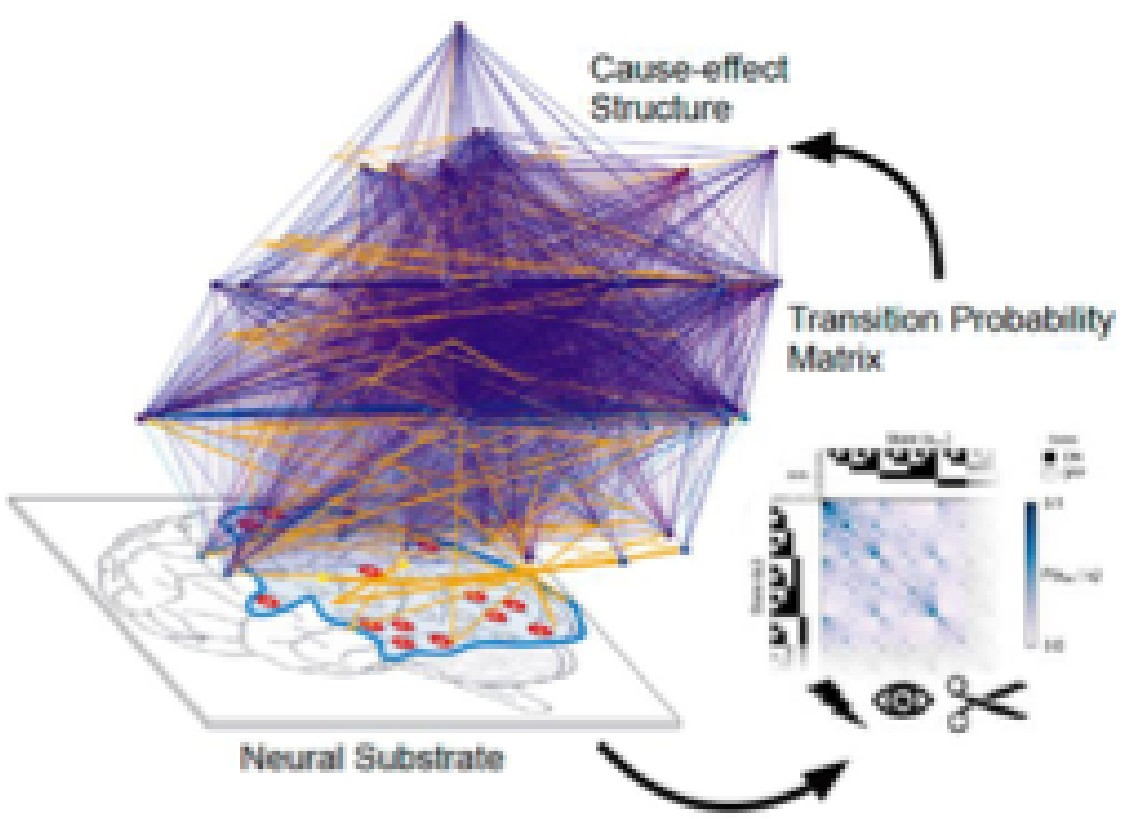

**IIT: Adversarial predictions**

**Exp.1:** Inactive neurons contribute to conscious experience

**Exp.2a:** Blindspot entails local shrinkage of perceived visual space

**Exp.2b:** Scotoma entails local shrinkage of perceived visual space

**Exp.3:** Active sampling not necessary for changes in conscious content

## Neurorepresentationalism

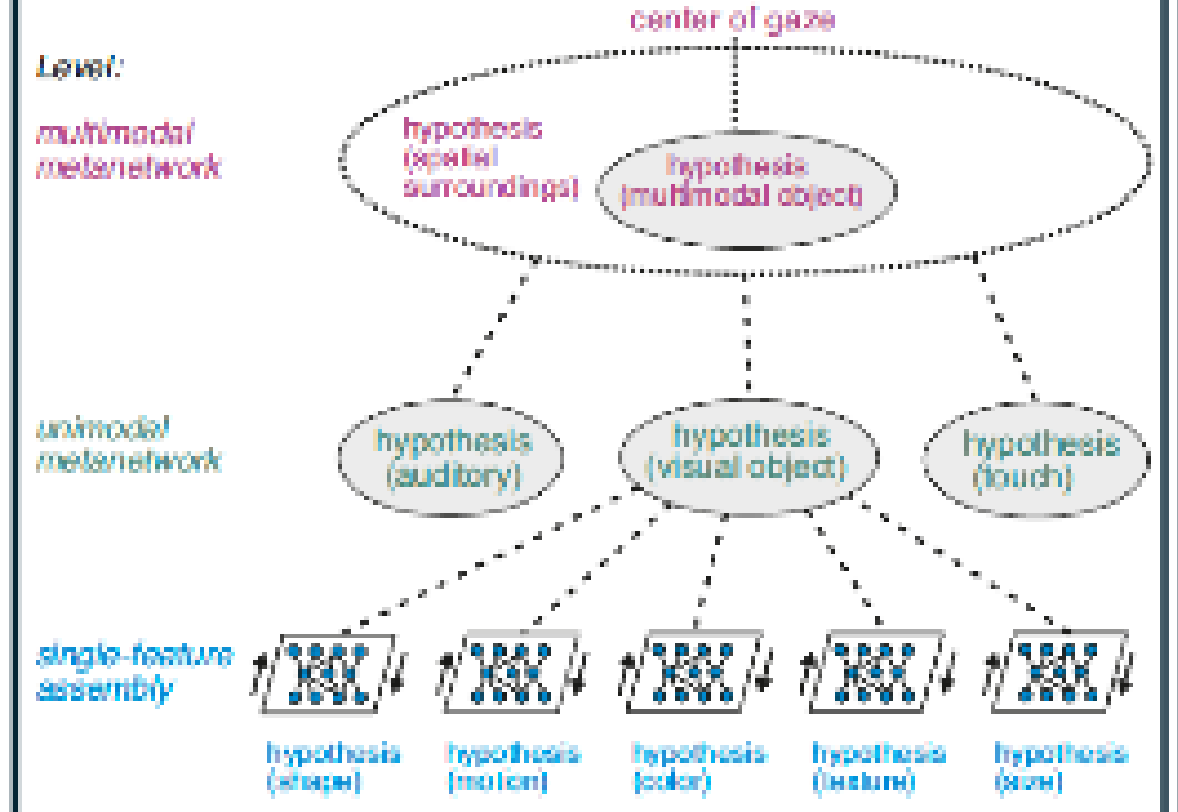

**NREP: Adversarial predictions**

**Exp.1:** Inactive neurons do not contribute to conscious experience

**Exp.2a:** Blindspot entails little or no shrinkage of perceived visual space

**Exp.2b:** Scotoma entails little or no shrinkage of perceived visual space

**Exp.3:** Active sampling not necessary for changes in conscious content

## Active Inference

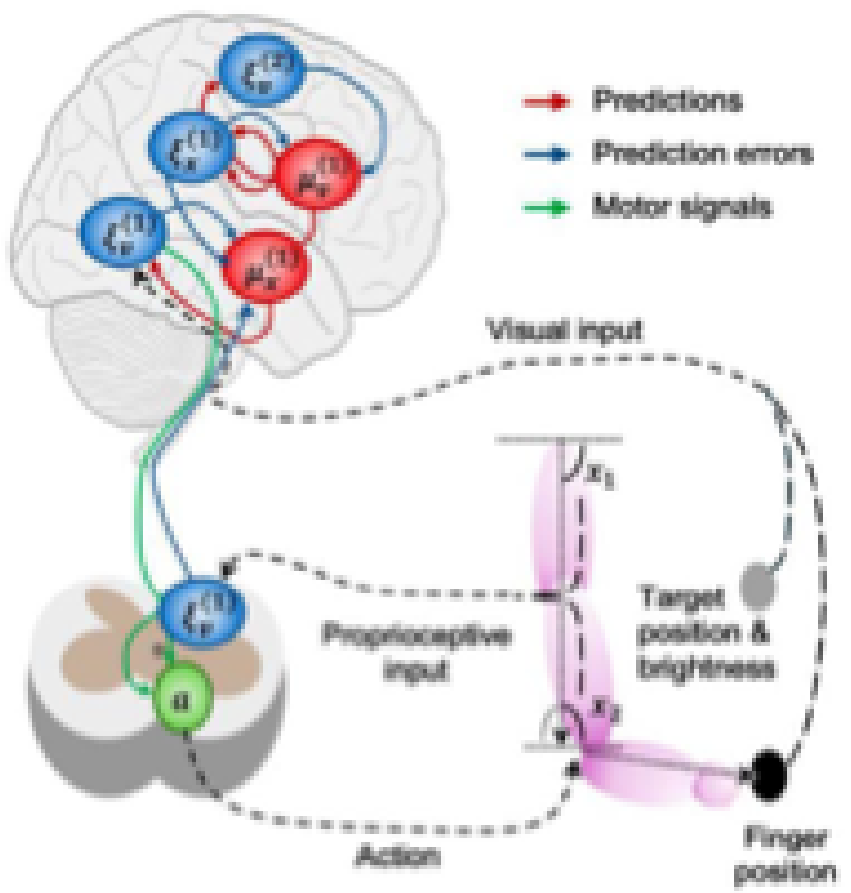

**AI-C: Adversarial predictions**

**Exp.1:** Inactive neurons do not contribute to conscious experience

**Exp.2a:** Blindspot entails no shrinkage of perceived visual space

**Exp.2b:** Scotoma entails no shrinkage of perceived visual space

**Exp.3:** Active sampling necessary for changes in conscious content

**Figure 1. The INTREPID project: An adversarial collaboration testing the Integrated Information Theory and two predictive processing theories of consciousness.** This schematic presents a high-level overview of the three theories of consciousness represented within the INTREPID Consortium, and the contrasting predictions to be tested during the adversarial collaboration. The core commitments of each theory are outlined in Section 2 of the main text; hypotheses and predictions are unpacked in Section 3. Panel illustrations adapted from Storm and colleagues (2024), Pennartz (2022), and Friston (2010).

# 2 Introducing the three theories

In this section, we begin by providing a very brief introduction to the core components of each of the three theories represented within the INTREPID Consortium, referring to more in-depth treatments along the way. We then highlight salient similarities and differences amongst the theories, focusing in particular on (1) the phenomena they seek to explain (*explananda*), (2) the sorts of explanation they proffer (*explanans*), and (3) the methodological approaches used to develop and validate their claims.

## 2.1 Key tenets of each theory

### 2.1.1 Integrated Information Theory (IIT)

IIT originates from the hypothesis that conscious systems evince high degrees of functional integration and differentiation, given the characteristically unified yet complex nature of subjective experience (Tononi & Edelman, 1998). Since its inauguration more than two decades ago (Tononi, 2004), IIT has undergone a number of refinements and elaborations (Albantakis et al., 2023; Balduzzi & Tononi, 2008, 2009; Oizumi et al., 2014; Tononi, 2008, 2012); however, the key idea that consciousness comprises fully-integrated, uniquely-specified states has remained at its core.

IIT begins by identifying the fundamental properties of consciousness – i.e., “those [properties] that are immediate and irrefutably true of every conceivable experience” (Albantakis et al., 2023, p. 2). These are distilled in the following set of ‘axioms’:

0. *Existence*. Experience *exists*; it is real.
1. *Intrinsicality*. Experience exists for *itself*; it is found from the experiencer’s own intrinsic perspective (independent of external observers).
2. *Information*. Experience is the particular way it is; it is *specific*, not indeterminate or generic.

3. *Integration*. Experience is *unitary*; it cannot be decomposed or reduced into separate parts without something being lost.
4. *Exclusion*. Experience is *definite*; its contents are bounded (a content is either included or not), and it unfolds at a particular spatiotemporal grain (neither finer nor coarser).
5. *Composition*. Experience is *structured*; it is composed of many interrelated phenomenological relations and distinctions that determine how it feels.

IIT then proposes a set of ‘postulates’ designed to explain how these phenomenal properties may be physically realised. Each postulate aims to characterise the specific causal properties a physical system would need to satisfy in order to support consciousness:

0. *Existence*. Physical existence is operationalised as *cause-effect power*; the physical substrate of consciousness must be able to ‘take and make a difference’.
1. *Intrinsicality*. The physical substrate of consciousness must have intrinsic cause-effect power; it must exert causal constraints on itself.
2. *Information*. The physical substrate of consciousness must specify a particular causal state (which may be measured in information theoretic terms).
3. *Integration*. The intrinsic causal power of the physical substrate of consciousness must be irreducible; parts of a partitioned substrate must, together, have less cause-effect power than the whole.
4. *Exclusion*. The intrinsic, integrated cause-effect power of the physical substrate of consciousness must be bounded; the bounds are those of the *maximal substrate* or *complex* (where any super/sub/set has less cause-effect power).
5. *Composition*. The intrinsic cause-effect power of the physical substrate of consciousness must be structured as a set of interrelated causal relations and distinctions; these are the individual mechanisms (groups of system units) that exert causes and effects within the substrate.

According to IIT, any entity that instantiates the physical properties specified by each of the postulates is conscious. The subjective quality of its experience depends on the form of the *cause-effect structure* or *Φ-structure* (“phi-structure”) derived from analysis (‘unfolding’) of the maximal substrate’s composition in a given state. The amount of consciousness expressed by the system is quantified by summing the integrated information $\varphi$ (“phi”) measured from each subset of units within the maximally irreducible cause-effect structure to calculate $\Phi$ (“structure-phi” or “big-phi”), which corresponds to the amount of *structure integrated information*. In this way, IIT provides a formal account of both the quality and the quantity of conscious experience (for technical details, see Albantakis et al., 2023).

Applied to the paradigmatic case of wakefulness in healthy human adults, IIT identifies temporal, parietal, and occipital regions of cerebral cortex – the so-called posterior 'hot zone' – as possessing the requisite organisational structure to instantiate consciousness in the brain, on the grounds that these regions exhibit dense, reciprocally interconnected and highly differentiated corticocortical architectures conducive to forming maximally irreducible cause-effect structures (in contrast to the more functionally modular and task-dependent connectivity attributed to much of frontal cortex; Tononi et al., 2016). The initial findings of another ARC adversarial collaboration comparing IIT against global neuronal workspace theory (Dehaene et al., 1998; Dehaene & Changeux, 2011; Dehaene & Naccache, 2001; Mashour et al., 2020) have been evaluated with respect to this claim (Cogitate Consortium et al., 2025; for discussion, see Negro, 2024), adding to pre-existing evidence from the clinical and neuroimaging literature (Boly et al., 2017; Koch et al., 2016; Storm et al., 2017; cf. Odegaard et al., 2017). The INTREPID Consortium aims to go one step further by testing the hypothesis that regions of this substrate specify the requisite cause-effect structure to explain both the quantity and quality of spatial experience (discussed in Sections 3.2 and 3.3).

### 2.1.2 Neurorepresentationalism (NREP)

NREP provides a predictive processing-inspired account of the way conscious experience arises from hierarchically-organised neural computations in the brain (Pennartz, 2015, 2022). Although the predictive processing framework offers a generic approach for understanding adaptive systems regardless of their conscious state (see, e.g., Clark, 2013, 2016; Friston, 2010; Hohwy, 2013, 2020; Piekarski, 2021), much work under the rubric of predictive processing is founded on predictive coding models of visual perception (Bastos et al., 2012; Bogacz, 2017; Lee & Mumford, 2003; Rao & Ballard, 1999; Srinivasan et al., 1982). NREP seeks to combine the fundamental principles enshrined in such models (namely, that the brain engages in a process of prediction error minimisation to infer the causes of sensory states) with insights drawn from philosophical analysis and neuroscientific data to elaborate a distinctive theory of consciousness.

NREP is rooted in the philosophical stance of 'representationalism', which aims to reconcile the subjective experience of elementary phenomenal properties (so-called 'qualia'; e.g., the redness of a tomato) with its material substrate – the brain (Pennartz, 2015, ch. 11). Qualia – and perceived properties in general – traditionally pose a problem for many materialist theories of mind insofar as these theories struggle to account for the physical location of such properties. For example, when we experience an illusion such as Kitaoka's rotating snakes (Murakami et al., 2006), it would be incorrect to state that the apparent rotational movement of snake-like shapes is physically happening in front of us. But it is equally incorrect to say that rotating snakes are literally present inside our brain (which consists of neurons, glial cells, blood vessels, etc.). So where is the quality of rotational motion to be found? Representationalism solves this problem by regarding perceived qualities as components of representations, which can be veridical or non-veridical (Lycan, 2023; Pennartz, 2018). The perceived rotation is therefore a represented property of the represented object – a property which in this case is illusory. The consequence of this account is that all of our conscious experiences are 'best-guess' reconstructions of the causes of sensory (or imagined) input – a notion going back to Kant (1787) and Von Helmholtz (1962).

It should be added that alternatives to representationalism exist, such as the disjunctive view of perception (akin to naive realism). Disjunctivism takes conscious perception to be veridical, successful presentation of the world; it directly presents environmental properties and brings the subject genuinely in contact with the world - in contrast to hallucinations and illusions which are not world-disclosing and are a fundamentally different kind of mental state (Soteriou, 2020). Although space is lacking to go into detail, NREP posits that direct perceptual access to the world is untenable because of the anatomical and physiological properties of the brain, which imply energy transduction in sensor cells, conduction of action potentials to brain systems, synaptic transmission (etc.), involving delays and time-consuming perceptual interpretive processes (Pennartz, 2018).

Similar to IIT, NREP deploys reflective analysis to first identify the 'inalienable features' distinguishing conscious experience from nonconscious processing in healthy human adults (Pennartz, 2015, ch. 8), which helps to demarcate what is understood by 'consciousness'. These 'hallmarks' of consciousness – which differ from IIT's axioms but show partial overlap with some other theories – are summarised as follows (Pennartz, 2022; Pennartz et al., 2019):

1. *Multimodal richness*. Conscious experience consists of distinctive sensory qualities derived from multiple modalities (e.g., vision, audition) and submodalities (e.g., colour, motion).
2. *Situatedness and immersion*. Conscious agents find themselves situated in a space that is organised with certain objects in the foreground and others in the background (context). One's body is experienced as immersed in the situation, occupying a central position relative to surroundings.

3. *Unity and integration*. Consciousness is characteristically unified such that the contents of experience form part of a singular, integrated whole (however, this does not imply that there is only one mechanism underlying integration).
4. *Dynamics and stability*. Static objects are experienced as stable elements in relation to moving objects and other changing inputs from the environment; both static and dynamic objects are distinguished from sensory changes brought about by bodily self-motion.
5. *Intentionality*. Conscious experience is about objects that differ from the neural substrates which instantiate them. Neural activity patterns lie at the basis of conscious experience; however, this experience is not about these activity patterns as such, but about the objects and situations the subject is aware of. These patterns are localised at different positions in space than where the neurons involved in the representation are located in the brain.

NREP interprets consciousness as furnishing the subject of experience with a unified, dynamic, multimodal 'survey' (or 'world-model') of unfolding bodily and environmental conditions (Pennartz, 2015, ch. 6). This inferential summary of the agent's current situation is argued to confer adaptive benefits, insofar as it underpins the ability to deliberate about and engage in goal-directed behaviour (Pennartz, 2018, 2022). Consciousness as conceived under NREP thus functions to inform on-the-fly, self-initiated decision-making and planned action (as opposed to reflexive or habitual modes of behaviour), but does not involve motor action *per se* (although it does involve 'activity' in the broader sense of generating top-down expectations, imagery, predictions of future sensory input, etc.). NREP thus acknowledges the importance of motor activity in shaping conscious experience, but does not consider such activity necessary for consciousness itself.

Some further qualifications of NREP need to be mentioned, as they clarify how NREP distinguishes itself from other theories of consciousness. To realise qualitative richness in conscious experience, NREP posits that interactions between unimodal and multimodal corticothalamic systems are required, both to segregate and integrate modalities (Pennartz, 2009). This integration takes place across multiple levels of representation, beginning with low-level predictions and errors concerning simple features (such as small, oriented edges in the visual field; i.e., local visual details), ascending to unimodal object representations (e.g., within the visual domain, but comprising different attributes such as colour, shape, texture, etc.), to finally reach a stage of very large, multimodal networks of networks (meta-networks) that perform a 'superinference' across multimodal inputs specified in space (Olcese et al., 2018; Pennartz, 2022). In this account, many forms of integration are posited to contribute to consciousness, including binding, grouping, binocular integration, scene synthesis by integrating across eye and head movements, and multisensory integration. For example, space perception critically hinges on the integration of retinotopic information with vestibular, proprioceptive, and other sensorimotor sources (e.g., efference copy), explaining how visual space perception transcends the level of retinotopic representation and thus realises craniotopic and allocentric aspects of perception (Pennartz, 2015, 2022, 2024).

### 2.1.3 Active Inference

Unlike IIT and NREP – which were both explicitly developed as scientific theories of consciousness – Active Inference explains the emergence of adaptive behaviour from 'first principles' (Parr et al., 2022). While this more general framework can be applied to conscious and nonconscious systems alike (Hohwy, 2022), an increasing number of researchers are drawing on its conceptual and methodological resources to explain various aspects of consciousness (e.g., Clark, 2019; Deane, 2021; Fields et al., 2025; Laukkonen et al., 2025; Rudrauf et al., 2017; Sandved-Smith et al., 2021; Seth & Tsakiris, 2018; Solms, 2019; Wiese, 2024; Williford et al., 2018; for helpful overviews, see Nikolova et al., 2022; Rorot, 2021; Vilas et al., 2022). We focus here on what has recently been dubbed 'the minimal theory of consciousness implicit in active inference' (Whyte et al., 2026), which is primarily concerned with the claim that active inference is necessary for a change in conscious content. We shall use the acronym AI-C to distinguish this narrower theoretical account from the broader (meta-theoretic) scope of the Active Inference framework at large.

Active Inference provides a normative account of sentient, self-organising behaviour (Friston et al., 2020; Parr et al., 2022; Sajid et al., 2021), where 'sentience' may be understood rather minimally as the expression of appropriate actions in response to sensory inputs (Friston, Da Costa, Sakthivadivel, et al., 2023; Friston et al., 2025). This perspective is inspired by the free energy principle (Friston, 2010; Friston, Da Costa, Sajid, et al., 2023), which provides a formalism for describing self-organisation — i.e., the emergence of adaptive agency — in random dynamical systems. Under this formalism, the internal states of the agent parameterise probabilistic beliefs about the external dynamics causing its sensory inputs under a generative or world model (Da Costa et al., 2021; Friston et al., 2020; Ramstead et al., 2023). Sentient behaviour on this view is characterised by the agent's capacity to model the expected consequences of its actions, thus enabling it to select the (approximately) Bayes-optimal course of action for a particular context (given the agent's prior preferences and beliefs; Friston et al., 2025; Hohwy, 2026; Pezzulo et al., 2024). This capacity gives rise to behaviour that appears planned towards an intended outcome (although not necessarily conscious). Mathematically, these inferential processes are underwritten by the minimisation of two objective functions: (1) variational free energy, which scores the degree of belief updating compelled by the observation of sensory data, and (2) expected free energy, which scores the probability of alternative courses of action (i.e., policies) based on the predicted outcomes those actions are expected to elicit.

In terms of computational and functional (biophysical) architectures – and attendant information theoretic characterisations — the free energy principle is compatible with most global brain theories, ranging from global neuronal workspace theories through to hierarchical predictive coding that is also used in NREP (e.g., Friston, Da Costa, Sakthivadivel, et al., 2023; Friston et al., 2012; Friston & Kiebel, 2009). However, the free energy principle — in and of itself — makes no claims about subjective experience.

Since the optimisation of variational and expected free energy is not claimed to be a specific feature or signature of subjective awareness, any theory of consciousness based on the Active Inference framework needs to explain how these two objective functions relate to conscious experience. Early work identified probabilistic beliefs encoded by the approximate posterior over hidden states as the best candidate for explaining conscious content (Hohwy, 2012; Hohwy et al., 2008), with the important caveat that only a privileged subset of posterior beliefs – namely, those spanning spatiotemporal scales relevant for embodied interaction with the world – contribute to conscious experience (Marchi & Hohwy, 2022; see also Clark, 2018). This proposal meshes well with the independently developed idea that subjective awareness is grounded in the 'temporal thickness' (or 'counterfactual depth') of the agent's generative model, which confers the capacity to select actions based on the inferred consequences of various possible choices (Friston, 2018; see also Clark et al., 2019; Corcoran et al., 2020; Friston et al., 2020, 2021; Hohwy, 2022, 2026).

After reviewing a variety of Active Inference models simulating the manipulation of conscious states and contents in synthetic agents, Whyte and colleagues (2026) proposed a minimal theory of consciousness (AI-C) reiterating the conceptual linkage between subjective experience and posterior beliefs about hidden states. On this account, all changes in conscious content – including the transition from unconscious to conscious experience and vice versa – must be driven by a change in the inferred state of the world (where 'world' here includes states within the brain and body, as well as the external environment). Moreover, AI-C identifies the interface between continuous (sensorimotor) and discrete (decision-making) levels of the processing hierarchy as the locus at which posterior beliefs become conscious, in line with previous work explaining conscious experience in terms of perceptual state inferences that function to inform policy selection (Hohwy, 2013; Marchi & Hohwy, 2022; Whyte, 2019). The upshot of this view is the claim that active inference is necessary (but may not be sufficient) for a change in conscious content (i.e., changes in consciousness must be accompanied by changes in posterior beliefs, but changes in posterior beliefs might occur without changes in awareness – e.g., belief-updates occurring at spatiotemporal scales too fast or slow to guide policy selection).

It is important to be clear that AI-C associates changes in consciousness with the deployment of a certain kind of inferential machinery to actively sample (or 'probe'; see Dołęga & Dewhurst, 2021) the sensorium – not the execution of an action *per se*. Under the Active Inference framework, actions are hidden states that must be inferred by the agent, and which are enacted as a consequence of some inferential process (e.g., the optimisation of expected free energy in the service of policy selection; cf. 'planning-as-inference'; Attias, 2003; Botvinick & Toussaint, 2012). It is also important to note that actions are conceived here rather generically, in a way that extends beyond the domain of motor control or voluntary behaviour. For instance, actions may pertain to cellular self-organisation, autonomic reflex arcs, covert attentional dynamics, or overt behaviours such as saccadic eye movements and subjective reports. What's common across these diverse phenomena is that they can all be understood as having been selected in accordance with beliefs about their sensory consequences, with a view to realising sensory states that ultimately minimise free energy. This means that AI-C can entertain changes in conscious experience that are driven by belief updates in the absence of overt movement – as in the case of changes in subjective awareness that are covertly entrained via the selection of attentional policies (i.e., 'mental actions'; Limanowski & Friston, 2018; Parr, Corcoran, et al., 2019).

## 2.2 Similarities and differences

Having briefly outlined the key claims of IIT, NREP, and AI-C, the remainder of this section presents a focussed comparison of their similarities and differences. We organise our discussion around three interrelated themes: (1) targets of explanation (*explananda*); (2) kinds of explanation (*explanans*); and (3) methodological strategies (see Table 1 for an overview of key comparisons).

### 2.2.1 Explananda: What do these theories attempt to explain?

Consciousness researchers disagree about the relevant phenomena any comprehensive theory of consciousness ought to explain (Francken et al., 2022; Seth & Bayne, 2022) – a situation Vilas and colleagues (2022) dub the 'explanandum problem'. This lack of consensus is perhaps somewhat surprising given our own intimate acquaintance with conscious experience (Chalmers, 1996; Wiese, 2018), but likely a product of vastly divergent intuitions about which features of mental life should be included within a scientific conception of consciousness (and how they ought to be operationalised for scientific investigation; Irvine, 2017; Phillips, 2018). In light of such diversity, it is important to be clear about the way consciousness is conceptualised under competing theories in order to disambiguate whether such theories make distinctive claims about the same target phenomenon, or whether they offer explanations targeting distinctive phenomena. While the latter case might be construed as a theoretical dispute about the fundamental nature of consciousness and the proper explananda that any scientific theory of consciousness science ought to target, such apparent disagreements might ultimately be resolved through subsumption under a broader, unifying theory of consciousness (Storm et al., 2024).

As discussed in Section 2.1, IIT, NREP, and AI-C all make explicit statements about their explanatory targets. IIT's axioms aim to provide an exhaustive list of the essential properties of consciousness, understood as the feeling of 'what it is like' to have an experience (Nagel, 1974). NREP's hallmarks function in a similar fashion, aiming to capture the universal features of healthy human conscious experience. Again, the main explanatory target here is the qualitative character of perceptual experience, with an emphasis on its situatedness and multimodal richness (Lee & Pennartz, 2025; Pennartz et al., 2019). While AI-C does not currently distinguish certain properties as essential or fundamental elements of consciousness (i.e., it does not posit 'axioms', 'hallmarks', or suchlike), its focus on the content of awareness indicates a similar concern with the phenomenal quality of experience. All three theories are thus broadly in agreement that a scientific theory of consciousness should account for the qualitative properties of experience – what is often referred to as 'phenomenal consciousness' (Block, 1995, 2011).

| | IIT | NREP | AI-C |
|---|---|---|---|
| **Explananda** | | | |
| Phenomenal consciousness | Axioms aim to provide an exhaustive list of the essential properties of subjective experience | Hallmarks aim to capture universal features of (healthy human) subjective experience | Aims to explain the subjective quality of experiential content; does not posit essential features |
| Access consciousness | Associates with cognitive and motor functions that operate on conscious states; not constitutive of consciousness | Associates with cognitive and motor functions that operate on conscious states; not constitutive of consciousness | Cognitive processes such as attentional regulation and decision-making may be constitutive of consciousness |
| States of consciousness | Aims to capture all possible states, which can be quantified by their amount of $\Phi$ and characterised by their $\Phi$-structure | Primarily targets wakefulness, but construes other states/modes (e.g. dreaming) as being graded along multiple dimensions | Primarily targets local states, but preliminary work indicates transition between global states can be modelled via changes in content |
| **Explanans** | | | |
| Conceptual | Characterises the qualitative content of experience in terms of the cause-effect structure of the physical substrate | Characterises content in terms of the multimodal integration of hierarchical neural representations | Characterises changes in content in terms of Bayesian belief updating and active sensory sampling |
| Mechanistic | Phenomenology fully explained by causal relations embedded in substrate; rejects functional or computational explanations | Low to medium level representations explained via predictive processing architecture; phenomenology achieved by high-level representations integrating multiple sensory modalities | Changes in conscious state and accompanying phenomenology fully captured by process theoretic explanations of precision-optimisation and policy inference |
| **Methodology** | | | |
| Identification of explananda | Axioms deduced from reflective analysis (‘phenomenology-first’) | Hallmarks derived from reflective analysis | Paradigmatic phenomena studied in consciousness science |
| Development and validation | Operationalise axioms via postulates, analyse cause-effect structure of candidate systems | Extend predictive coding models and examine neural substrates by recordings and causal perturbations | Construction and analysis of Active Inference models of paradigmatic phenomena |

**Table 1. Summary of similarities and differences in the explananda, explanans, and methodological approaches of Integrated Information Theory (IIT), Neurorepresentationalism (NREP), and the Active Inference theory of consciousness (AI-C).** See main text for further elaboration of these points.

Beyond phenomenal consciousness, Active Inference has previously been deployed to model features of 'access consciousness' (Whyte et al., 2022; Whyte & Smith, 2021) – those elements in awareness that can be reported, thought about, and used to guide behaviour (Block, 1995, 2011). Indeed, AI-C might prove particularly well-suited for characterising aspects of access consciousness, given the emphasis it places on such executive processes as attentional regulation, decision-making, and action planning. This may explain the historical affinity between Active Inference based accounts of consciousness and Global Workspace Theory (Hohwy, 2013; Whyte, 2019), which identifies access consciousness as the primary (and perhaps exclusive) explanandum for the science of consciousness (Cohen & Dennett, 2011; Dehaene et al., 2006; Naccache, 2018). In contrast, IIT and NREP both associate access consciousness with cognitive and motor functions that operate on phenomenally conscious content (rather than being constitutive of consciousness *per se*; Ellia et al., 2021; Pennartz, 2015). This highlights a difference between IIT and NREP on the one hand as theories that determine their explananda on pre-theoretical grounds (i.e., through reflective analysis of the essential features of conscious experience), versus AI-C on the other as a theory that targets those phenomena that are paradigmatically studied in consciousness science at large (see Section 2.2.3).

Leaving the distinction between access and phenomenal consciousness aside, one can also ask how each of the three theories considered here relates to states of consciousness (Chalmers, 1996). One way of categorising states of consciousness is to distinguish global states (pertaining to the overarching profile of subjective experience) from local states (pertaining to the specific contents of subjective experience; Seth & Bayne, 2022). From its inception, IIT has sought to address both the quantity and quality of experience (Tononi, 2004, 2008) – thus aiming to provide a comprehensive account of both the global and local properties of conscious states. On this view, global states (e.g., wakefulness, dreaming, coma) are interpreted as 'levels' along a unidimensional scale corresponding to $\Phi$, where any system with $\Phi > 0$ is conscious to some degree (note that calculation of $\Phi$ is not feasible beyond very simple networks, although a variety of proxy measures have been proposed; Mediano et al., 2019, 2022). The qualitative character of conscious states is inferred from the form of their substrate's unfolded cause-effect structure.

NREP and AI-C are not committed to the view that global states can be straightforwardly quantified and ordered along a single scale as implied by IIT (cf. Bayne et al., 2016). In NREP, global states are associated with modes of consciousness (perception, imagery, dreaming) or states of low or absent consciousness (e.g., deep sleep, anaesthesia, coma). AI-C does not really admit the notion of a conscious 'state'. It considers conscious (and unconscious) inference to the best explanation – and planning as inference – as 'processes'. Insofar as NREP takes consciousness to subserve goal-directed behaviour, it focuses in the first instance on explaining the neural underpinnings of perceptual content experienced in the context of wakefulness (since these are the states most relevant for prospection and goal-directed behaviour, which NREP construes as the targets subserved by consciousness). However, NREP also aims to account for other forms of conscious experience that are less dependent on external sensory processing, as in dream states and mental imagery (Pennartz et al., 2019). For instance, dreaming is conceived of as a 'virtual reality' state characterised by being less susceptible to cognitive control and less strongly coupled to output systems for goal-directed behaviour (Pennartz, 2015). It is not however considered to be a 'lower' state of consciousness than wakefulness, but rather to constitute a different mode. Nonetheless, NREP recognizes that consciousness can be graded along several dimensions – intensity, spatiotemporal resolution, and multimodal richness.

Although AI-C is principally concerned with local states insofar as it aims to account for changes in the content of experience, its scope could conceivably be broadened to capture different sorts of global states. Indeed, the transition from an unconscious to a conscious state may be construed as a change from a state with no content to one with some content (Whyte et al., 2026). Whyte and colleagues extend a previously reported hierarchical model of auditory processing (Smith et al., 2022) to capture differences in evoked neural responses to local and global regularities which have been empirically documented across sleep and wake states. This model suggests that the loss of consciousness in deep sleep is driven by the disconnection of hierarchical levels within the generative model, resulting in a breakdown of message-passing that precludes belief-updating within temporally-deep levels. This example illustrates how existing Active Inference models of sensory processing might be elaborated to account for differences in the phenomenology and neural dynamics associated with various global states of consciousness.

### 2.2.2 Explanans: How do these theories explain consciousness?

Having established the relevant explananda of IIT, NREP, and AI-C, we turn now to their respective *explanans* – the kinds of explanation each theory proffers for their explanatory targets. Again, we caution that differences in theoretical explanations of common explananda may not necessarily entail a fundamental disagreement – one theory's explanans may be compatible with (or even equivalent to) another's. This being said, one explanatory approach may still be preferred over another, equally viable explanation for a variety of reasons (e.g., theoretical virtues such as simplicity, fecundity, unification, etc.; Kuhn, 1977; Quine, 1955). We focus here on the kinds of explanation availed by each theory, deferring treatment of their implications for empirical testing to Section 3.

IIT proposes an explanatory identity between conscious experience and the cause-effect structure 'unfolded' from its physical substrate (Albantakis et al., 2023; for discussion, see Cea et al., 2023). This implies an isomorphic mapping between the phenomenal features of a given experience and the causal configuration of the physical substrate instantiating it – that is, the subjective quality of experience is fully explained by the cause-effect powers observed in the conscious system. IIT thus renders explanations of conscious phenomena expressed in terms of the causal relations embedded within their substrates. An important implication of this approach is its rejection of functionalism: two systems may have equivalent (or practically indistinguishable) input-output functions (i.e., respond to identical stimuli in the same way), yet the degree and quality of conscious experience may differ dramatically between systems depending on the cause-effect powers realised by their architectures (Albantakis et al., 2023; Grasso et al., 2021; Oizumi et al., 2014). For this reason, IIT does not regard functional (or behavioural) properties as reliable indicators or satisfying explanations of consciousness (Ellia et al., 2021; Tononi & Koch, 2015; cf. Cohen & Dennett, 2011).

In contrast to IIT's focus on intrinsic causal structure, NREP and AI-C ground their explanations of consciousness in information processing architectures that support inferences (i.e., probabilistic beliefs) about external states of affairs. NREP claims conscious experience is the product of 'superinference' on the causes of sensory input in which all levels of the multimodal processing hierarchy participate (Pennartz, 2015, ch. 10, 2018, 2022). Although NREP subscribes to the predictive processing framework, it is not wedded to any specific process-theoretic description of the computational architecture underpinning such inferences (see, e.g., Brucklacher, Lee, et al., 2025; Salvatori et al., 2021; Spratling, 2017; Sprevak & Smith, 2023). NREP has provided computational network models based on sensory cortical hierarchies, which illustrate several integrative processes operating at the level of specific image representation, view-invariant object representation, figure-ground segmentation during ego- and object-motion, and predictive cross-modal interactions (Brucklacher et al., 2023; Brucklacher, Pezzulo, et al., 2025; Dora et al., 2021; Pearson et al., 2021). These models provide mechanistic explanations of the low-to-medium levels of representation (i.e., of single features and unimodal objects); however, NREP abstains from attempting to precisely equate or map computational models to phenomenology, since this transition from quantitative to qualitative properties is deemed unimaginable (Lee & Pennartz, 2025; Pennartz, 2015).

AI-C can similarly be described as presenting a predictive processing account of conscious experience; however, unlike NREP, it is specifically committed to the process theory developed under the free energy principle (Friston et al., 2017; Parr et al., 2022) – with the caveat that the precise implementational details described at the process theory level are themselves subject to development (e.g., there are multiple message passing schemes consistent with the free energy principle that could be used to implement an Active Inference model; Parr, Markovic, et al., 2019). While AI-C proposes that the inferential (Bayesian) mechanics underpinning changes in conscious contents must conform to the free energy principle – and thus adduces the optimisation of free energy functionals as its primary explanans – NREP emphasises the more classic and specific concept of prediction error minimization (cf. Rao & Ballard, 1999). Instead, NREP posits that basic predictive coding models take care of low-level computational operations, but do not present the appropriate descriptors for phenomenology playing out at the highest representational levels.

The crucial role of policy selection in AI-C highlights a distinctive explanatory construct that is absent from both IIT and NREP (and that is the focus of one set of experiments being conducted by the INTREPID Consortium; see Section 3.4). As mentioned in Section 2.1.3, active inference pertains to the optimisation of beliefs about precisions and policies; there is no requirement for overt movement to be initiated in order for changes in awareness to occur (consonant with IIT and NREP, both of which dissociate overt action from conscious experience). However, the notion of mental action – in which covert attentional policies are selected in a way analogous to those driving overt behaviour (Limanowski & Friston, 2018; Parr, Corcoran, et al., 2019) – highlights the fundamental explanatory role of such executive processes as attentional regulation (i.e., precision-optimisation) and decision-making (i.e., planning-as-inference) in AI-C. This formulation is reminiscent of certain action-oriented or enactive approaches to consciousness, such as sensorimotor contingency theory (O'Regan, 2011; O'Regan & Noë, 2001; see also Seth, 2014). Arguably, AI-C benefits from a formalised approach that grounds the "everyday-language" explanations afforded by sensorimotor theory (O'Regan, 2023, p. 2) – e.g., the notion of being "poised" to act in a particular way (O'Regan, 2022, p. 3) – in a computationally tractable account that speaks to underlying neural dynamics (Nave et al., 2022). In short, NREP considers representation sufficient for 'seeing', while AI-C posits that representation is necessary but not sufficient, in the enactivist sense that 'to see is to look' (overtly or covertly).

### 2.2.3 Methodology: How are these theories constructed and validated?

Finally, we consider the methodological approaches adopted to develop and validate each of the theories addressed in this review. We briefly outline both the strategies used to construct or derive core elements of each theory, and how these strategies guide the empirical investigation of consciousness under each theoretical perspective.

IIT distinguishes itself from most other neuroscientific theories by adopting a 'phenomenology-first' approach that seeks to characterise the intrinsic structure of consciousness (Ellia et al., 2021; Negro, 2020; Oizumi et al., 2014). Beginning with introspection and reasoning, the essential, invariant properties of subjective experience are derived to form the axiomatic foundations of the theory. These phenomenological properties are then operationalised in terms of physical (causal) properties (i.e., postulates) that are amenable to formal analysis (e.g., by quantifying the amount of integrated information expressed by a causal network) and empirical investigation (e.g., by searching for neural substrates capable of supporting particular cause-effect powers). The operationalisation of axioms into their corresponding postulates is based on an 'inference to a good explanation' under the assumption there exists an observer-independent reality comprising reliable causal relations that can be analysed in terms of their smallest constituents (Albantakis et al., 2023; Chis-Ciure, 2022; Tononi et al., 2022; cf. Cea et al., 2023).

The mathematical framework provided by IIT enables one to quantify and evaluate the various properties specified by the postulates for a candidate system, culminating in a formal description of the qualitative structure of experience under a given state. In order to validate this formalism, one may compare the kind of structure it prescribes for certain kinds of conscious experience with the physical organisation of neural structures involved in supporting such experience. For example, an IIT-based analysis of spatial extendedness suggests a grid-like substrate would be required to instantiate this kind of experience, which is consistent with the structure of posterior cortical regions implicated in visual processing (Grasso et al., 2021; Haun & Tononi, 2019). This observation leads to the hypothesis that perturbations of this grid-like neural structure should distort the phenomenal quality of visual space in a predictable manner (Haun & Tononi, 2019; Tononi et al., 2016; see Section 3.3). Analysis of the sort of structure instantiated by the cerebellum, on the other hand, is argued to confirm its poor suitability for supporting consciousness (due to low levels of neural integration; Tononi, 2008; Tononi et al., 2016). Once this formalism has been sufficiently validated across various states of consciousness in healthy human adults, it may then be extended to characterise the quality and quantity of experience in more challenging cases (e.g., infants, unresponsive patients, non-human systems; Tononi & Koch, 2015).

Although NREP does not explicitly specify a single favoured strategy for theory development and validation, it blends aspects of the philosophical stance of representationalism with empirical and computational methods from neuroscience and robotics. NREP-inspired research aims to complement the search for neural correlates of consciousness (Crick & Koch, 1990, 2003) by investigating the impact of causal (e.g., optogenetic) manipulations of neural substrates, and by developing predictive coding-inspired computational models capturing the network dynamics hypothesised to underpin representational processes. Following this strategy, researchers have demonstrated, for instance, the neurobiological plausibility of predictive processing-based representational models in spiking neural networks (Lee et al., 2024), and reproduced the increase in population sparsity and single-neuron image selectivity observed when ascending the visual cortical hierarchy (Dora et al., 2021).

Notably, NREP's primary objective is to characterise the neural basis of representations pertaining to externally- and internally-driven modes of experience as instantiated in healthy human adults (Pennartz, 2018, 2022). While this objective may appear to hinder the potential expansion of NREP's explanatory scope to include other animal species and artificial agents, this need not be the case. Indeed, it may be possible to abstract the computational architecture underpinning human consciousness away from its neural implementation and apply these insights in other domains (see, e.g., Pennartz, 2015, ch. 11; Pennartz et al., 2019). For instance, the effectiveness of multisensory integration in predictive coding driving place recognition in mobile rodent-like robots (Pearson et al., 2021) exemplifies how this approach may help to define so-called 'indicators of consciousness' in artificial agents (Lee & Pennartz, 2025; Pennartz et al., 2019).

In contrast to IIT's 'phenomenology-first' approach and NREP's unique blend of philosophical, empirical, and computational analysis, AI-C might be characterised as adopting a 'model-first' approach whereby core elements of AI-C are derived from the interrogation of process-theoretic models. This approach is inspired by Hohwy and Seth's (2020) proposal to deploy insights from predictive processing in the service of consciousness research and theory development. The key idea here is to exploit the resources availed by a domain-general modelling framework such as Active Inference to construct computational models of paradigmatic phenomena in consciousness science (e.g., binocular rivalry; Doerig et al., 2021), and to compare the properties of these models such that salient commonalities and particularities may be identified (Vilas et al., 2022; Whyte et al., 2026). Given a sufficiently diverse set of models, the hope is that systematic examination of their properties will eventually yield meaningful insights into the computational differences between conscious and unconscious states, and the content of those states (one might conceive of this endeavour as a search for 'computational correlates of consciousness'; Cleeremans, 2005; Wiese & Friston, 2021).

A distinctive feature of the model-first approach used to develop AI-C is that it may help to minimise the potential influence of background assumptions that constrain both the phenomena targeted for explanation and the kinds of observation deemed relevant for theory development and evaluation (cf. Yaron et al., 2022). Rather, a wide variety of empirical phenomena and methodological paradigms can be exploited in pursuit of the computational properties of consciousness, independent of prior beliefs or assumptions about the nature of consciousness itself (as derived, e.g., via introspective reflection). While it is currently unclear how well this strategy will perform when applied to more challenging or controversial cases (e.g., artificial systems), the delineation of common computational properties across a range of paradigmatic cases may provide a useful starting point for inferring the existence and contents of conscious states within such systems.

# 3 From theoretical constructs to empirical predictions

Having sketched out the key tenets of IIT, NREP, and AI-C – as well as some notable similarities and differences between them – we turn next to the topic of their empirical validation. The overarching premise of the Accelerating Research on Consciousness (ARC) initiative is that theoretical disagreements about the nature of consciousness may be settled by devising and implementing experiments that test distinctive hypotheses derived from each theory (Melloni et al., 2021; Reardon, 2019). Inspired by the historical example of Sir Arthur Eddington's seminal *experimentum crucis* – in which competing predictions of Newtonian mechanics and general relativity were pitted against one another on the occasion of the 1919 solar eclipse (Dyson et al., 1920) – such experiments are designed to generate data that will corroborate the predictions of one theory while simultaneously disconfirming those of its competitors (Del Pin et al., 2021; Negro, 2024). Although the findings of a few experiments are seldom sufficient to definitively refute an entire theory – perhaps no bad thing considering the relative immaturity of consciousness science (Evers et al., 2024; Negro et al., 2024; Wiese, 2018) – such data are expected to inform the debate (and future research) by clarifying which theory enjoys the most empirical support (Corcoran et al., 2023).

In this section, we begin by considering some of the important steps and challenges that complicate attempts to test scientific theories, and the role of adversarial collaboration in addressing (or at least mitigating) some of these issues. We then turn our attention to three key hypotheses that will be investigated by the INTREPID Consortium. Our goal here is to succinctly outline the theoretical motivation for each hypothesis and to anticipate the potential implications of alternative empirical outcomes for each theory.

## 3.1 Bridging the gap between theory and observation

Before sketching out the distinctive hypotheses that will be tested by the INTREPID Consortium, it is important to consider the inherent challenges posed by theory testing in general – as well as those confronting neuroscientific theories of consciousness in particular. On the first point, we note that theories generally consist of a set of more-or-less formalised propositions that are not themselves amenable to direct empirical validation (e.g., the axioms and postulates of IIT; conformity to the free energy principle). Rather, a number of additional concepts and assumptions must be invoked to bridge the gap between the theoretical and empirical realms (e.g., to compute the integrated information of a network of units; to describe how predictive processing mechanisms are implemented in the brain). This generally involves translating a set of theoretical propositions or principles into a formal process theory that in turn generates predictions about target phenomena that may be subjected to empirical validation (Devezer & Buzbas, 2023; for a philosophical overview, see Vorms, 2018). If predictions are borne out by the data, this lends support to the theory from which they were derived (at least to the extent the process theory embodies the relevant theoretical constructs; see Negro et al., 2024). However, if predictions fail to adequately capture the data, it is unclear whether the fault lies within the core of the theory itself, or within the additional set of auxiliary hypotheses and background beliefs that were enlisted to render the theory empirically tractable (Duhem, 1954; Quine, 1951). Such ambiguity enables the resourceful theoretician to preserve their core theoretical commitments at the expense of peripheral factors which can be cheaply sacrificed in the face of challenging data (e.g., changing the mathematical formalism used to calculate integrated information; updating the message passing scheme used to implement predictive coding or active inference) – hence why such auxiliary components are sometimes said to form a ‘protective belt’ around the core elements of one’s theory (Lakatos, 1976).

Given the nascent state of consciousness science – and the scientifically unusual situation of seeking reliable empirical data about the quality of subjective experience – the gap between theory and observation is often considerable. Theories of consciousness are in general not formalised to the degree that predictions about target phenomena can be simply and straightforwardly derived (i.e., in the way that Newtonian mechanics and general relativity both entail precise predictions about the degree of gravitational lensing that ought to be observed during a solar eclipse). The implications of a theory of consciousness for specific phenomena may be ambiguous or underdetermined (see, e.g., Bayne, 2018). Moreover, as discussed in Section 2.2, there is little consensus about the phenomena that ought to be targeted for explanation by scientific theories of consciousness – or the methodological procedures that ought to be deployed to study them (see also Mudrik et al., 2025). This is akin to followers of Newton and Einstein disagreeing about the relevance of light’s apparent deflection around celestial objects for explanations of gravity, or disputing whether telescopes are an appropriate instrument for measuring the degree of gravitational lensing caused by the sun.

Adversarial collaboration does not eliminate such fundamental difficulties. It does however provide a mechanism for ensuring that the data to be collected in a planned study are agreed to be informative for the dispute at hand, meaning its results will need to be taken seriously by adherents of competing theoretical perspectives irrespective of the outcome. This does not mean that co-designed experiments should be expected to definitively arbitrate amongst rival views; one paradigm might prove more critical for one particular theory, insofar as it tests predictions that lie closer to its core set of principles and propositions than another’s (Negro, 2024). Neither does it prevent theoreticians from ascribing unanticipated or unfavourable outcomes to faults within their auxiliary hypotheses or background assumptions (rather than their theory’s core) – indeed, this is perfectly rational behaviour which may precipitate important theoretical developments and empirical discoveries (Chalmers, 2013; Corcoran et al., 2023; Lakatos, 1976). But collaborators should agree from the outset that the questions to be addressed by their co-designed experiments are capable in principle of generating the sort of data needed to progress the debate, and that the methods used to obtain and analyse such data are sufficiently rigorous and reliable to obviate facile dismissal.

These considerations led the INTREPID Consortium to develop a multi-experiment project designed to jointly test unique combinations of predictions elicited from each of the three theories under examination. In this approach, no single experiment is decisive for arbitrating disagreements amongst proponents of IIT, NREP, and AI-C. However, certain experiments are more heavily weighted towards testing a key prediction of one theory versus its competitors; hence, when aggregated together, the outcome of these experiments will indicate the relative performance of each theory across a variety of settings. In the remainder of this section, we adumbrate the key hypotheses to be tested in this project, and briefly illustrate how they may be related to core and auxiliary components of each theory. For more detailed exposition of the full range of predictions, paradigms, and analysis plans for each experiment, please refer to the Consortium's preregistration document (Olcese et al., 2024) and the experiment-specific Study Protocols (Abbatecola et al., 2026; Haun et al., 2026; Robinson et al., 2025; Takahashi et al., 2025).

## 3.2 Hypothesis #1: On the contribution of inactive neurons to the specification of visuospatial awareness

According to IIT, the quality of any conscious experience is determined by the form of the cause-effect structure corresponding to that experience. Since such structures are specified by the cause-effect powers inherent within the organisation of their physical substrate, any alteration of its architecture should be accompanied by a corresponding alteration of experience. An intriguing consequence of this view is the claim that consciousness may be sustained by a cerebral cortex which is *silent* in terms of neural activity (Oizumi et al., 2014; Tononi et al., 2016). This follows from the premise that silent (i.e., quiescent or 'inactive') neurons are just as important for specifying the cause-effect structure as are active neurons, since silent neurons contribute to the exclusion of alternative neural configurations that would otherwise have given rise to different conscious experiences. IIT thus posits that intervening on inactive neurons such that their functional capacity for activation is disabled – i.e., their potential ability to 'take or make a difference' is abolished – should alter subjective experience on account of preventing these neurons from participating in the cause-effect structure (Albantakis et al., 2023).

The notion that conscious experience may be manipulated by suppressing the potential activation of already inactive neurons is a bold and perhaps counterintuitive claim – one that distinguishes IIT from many other theories of consciousness, including NREP and AI-C. While the latter two theories do not make positive claims about the specific role of inactive neurons in awareness, the inactivation of already-inactive neurons should not be expected to exert any significant effect on conscious experience under these accounts. This is because inactive neurons are not expected to influence the rest of the network they are a part of; hence, their 'inactivation' (where their capacity for activation is suppressed) should be indistinguishable from their mere quiescence (where their capacity for activation remains intact). In other words, an experimental manipulation that does not perturb neural dynamics should not be expected to perturb subjective experience.

There is, however, an important distinction between NREP and AI-C that leads to divergent predictions about the contribution of 'background neuronal activity' to spatial awareness. Because NREP claims that consciousness arises from a multimodal, integrative 'superinference' on representations including object relations (Pennartz, 2009, 2015), this implies that some degree of background neuronal activity is necessary for generating consciousness (Pennartz, 2022; Pennartz et al., 2023). Hence, although the 'inactivation' of already inactive neurons is not anticipated to affect conscious experience, suppression of background activity is expected to have a disruptive effect (providing this activity is strong enough to impact other neurons; Pennartz, 2015, 2022). AI-C does not entail this prediction, although it is plausible that the suppression of background activity could alter precision estimation in ways that impact visuospatial awareness and behaviour (Friston, 2018; Parr & Friston, 2018).

Optogenetic techniques enabling the precise manipulation of neuronal activity afford the tantalising prospect of investigating the impact of neuronal inactivation on visuospatial perception in rodents and other animals. The INTREPID Consortium aims to exploit such methodological advances in the service of testing the contrasting predictions of IIT, NREP, and AI-C. Briefly, rodents will be trained to perform a discrimination task in which they must indicate whether a target stimulus presented in the left and right fields of view falls within the left or right (subjectively perceived) hemifield. In critical trials, regions of visual cortex in the left hemisphere will be optogenetically hyperpolarised to investigate the impact of transient bouts of neural inactivation on task performance. If silent neurons (or those evincing background levels of activity) are indeed necessary for the conscious experience of space, inactivating these populations should precipitate systematic changes in behavioural performance that are consistent with perturbations of spatial awareness akin to hemineglect or hemianopia (for further details, see Takahashi et al., 2025).

As is typically the case with sophisticated research designs, this experiment presents a number of technical challenges that must be overcome if it is to generate meaningful data.

First, the animals must be capable of performing the task well enough such that (changes in) spatial awareness can be reliably inferred from (changes in) their behaviour. Second, inactive neurons must be reliably distinguished from active neurons, and the experimental manipulation must be implemented with sufficient precision such that only the target subset of neurons are inactivated as intended. This task is complicated by the fact that healthy neurons typically express a certain degree of spontaneous activity, meaning that subsets of inactive neurons may need to be classed according to a threshold level of quiescence that does not correspond to absolute silence. Third, it must be ensured that the experimental manipulation does not induce unintended side effects on downstream processes that might confound the effect of neural inactivation on consciousness *per se* with (for example) decision-making and behavioural processes bearing on the dependent variables being used to infer (changes in) spatial awareness.

These three exemplary challenges serve to illustrate some of the methodological difficulties involved in designing informative experiments capable of arbitrating theoretical disputes. Failure to select an appropriate task for tracking changes in conscious experience would undermine the experiment from the get-go, since the data generated through such procedures may not pertain to the desired explanandum. Failure to accurately identify and precisely manipulate neural activity may produce changes in conscious experience that can be reliably measured but cannot be ascribed to the causal mechanism hypothesised under the theories in question. The inferences licenced by a set of empirical observations thus depend on a conjunction of hypotheses that pertain not only to core theoretical claims of interest, but also to auxiliary assumptions about the adequacy of the interventions and measures used to probe these claims. Importantly, adversarial collaboration may help to reduce disagreements about certain auxiliary assumptions (e.g., about whether an experimental paradigm is appropriate for testing hypotheses about conscious experience, what level of activity constitutes a reasonable threshold for classing neurons as inactive, etc.). However, the space of potential auxiliary hypotheses that may be implicitly tested by any given experimental procedure is vast and seldom fully constrained prior to data collection (Lakatos, 1976).

Before moving to the next hypothesis, it is worth noting that auxiliary hypotheses are not exclusively concerned with methodological matters such as those identified in the examples enumerated above. For example, assume that the optogenetic manipulation designed to test the effect of inactivating neuronal circuitry is executed perfectly, yet no difference in task performance is detected. Moreover, assume control manipulations reveal that the rodents respond to task stimuli as expected, indicating that their behaviour yields a reliable indicator of their subjective experience. This outcome would constitute evidence against IIT's prediction about the effect of neuronal inactivation, but it would not refute the more fundamental idea that intervening on the physical substrate of the cause-effect structure in a way that changes its repertoire of potential states will necessarily correspond to a change in subjective awareness. This is because cause-effect structures are not assumed to be specified at the neuronal level; manipulations of neuronal activity may not constitute interventions on the correct spatiotemporal grain to influence the substrate's cause-effect powers. Thus, IIT could accommodate the failure to observe any effect of neuronal inactivation, at the cost of finding a plausible alternative candidate substrate. That is, new auxiliary hypotheses would be required to translate core principles of the theory into empirically-tractable predictions.

## 3.3 Hypothesis #2: On the contribution of cortical structure to the quality of spatial extendedness

A further implication of the identity IIT posits between the quality of experience and the cause-effect structure is that fundamental properties of experience can be explained in terms of the composition of the structure as specified by its physical substrate. Proponents of IIT have identified the 2D grid-like organisation of posterior cortical networks as a prime candidate for explaining the phenomenal character of spatial extendedness in the visual modality (e.g., the subjectively perceived distance between two points in the visual field; Grasso et al., 2021; Haun & Tononi, 2019). IIT predicts that changes in the connectivity of primary visual cortex should cause systematic alterations in the way space is experienced (Albantakis et al., 2023; Haun & Tononi, 2019). For example, lesions of primary visual cortex are expected to result in a corresponding collapse of the experience of spatial extent, such that objects spanning these regions appear closer together than they otherwise would if the intervening cortical structure were intact (Haun & Tononi, 2019).

NREP and AI, by contrast, do not predict gross distortions of spatial experience as a consequence of such structural alterations or lesions. Rather, predictive processing mechanisms endorsed by these theories imply that regions of the visual field corresponding to such areas should instead be 'brushed-over' or 'filled in' (as has been reported for artificially-induced perceptual scotomas; Ramachandran & Gregory, 1991). Since NREP argues that spatial experience arises from a large-scale inferential process involving multiple modalities and underlying brain systems (e.g., visual input, vestibular and proprioceptive inputs, etc.), estimates of spatial distances spanning damaged or missing cortex should average out with estimates from trajectories through intact portions of the visual field. While AI-C predicts that spatial estimates may be less accurate — due to loss of precise sensory evidence — it predicts no systematic bias corresponding to a contraction or dilation of perceived space (other than a regression to prior expectations).

The INTREPID Consortium aims to test these competing predictions through two complementary experiments – one which treats the retinal blindspot as a naturally-occurring lacuna within cortical space (Abbatecola et al., 2026), and one examining the effect of neurological insults resulting in a scotoma (Haun et al., 2026). The motivation for both experiments is the same – to investigate whether target stimuli spanning or bordering the blindspot/scotoma feel closer together (or further apart) compared to foils presented elsewhere in the visual field. IIT predicts that the absence of functional neural tissue in these regions of cortex translates to an altered cause-effect structure in which the region of space that would have been represented as part of the visual scene is simply missing – hence making target stimuli appear closer together (since there is less intervening phenomenal space between stimuli). NREP predicts no (or only small) distortions of spatial extendedness; AI-C predicts no bias but reduced accuracy.

IIT's prediction of subjective spatial contraction is concordant with psychophysical evidence from healthy adults (Song et al., 2017); however, other forms of distortion (e.g., spatial dilation) may also be accommodated by the theory according to the analysis presented by Haun and Tononi (2019). One lack of constraint pertaining to the nature of the anticipated distortion concerns uncertainty over the precise structural connectivity evinced by the hypothesised grid-like networks in primary visual cortex – a problem that may be exacerbated by individual differences in the microstructure of these regions. It is also unclear whether adaptive compensatory mechanisms, acting from birth onwards, may affect the way visual experience is constructed for regions of space corresponding to the blindspot. Moreover, testing this hypothesis in the context of scotoma introduces additional challenges; psychophysical experiments with many trials are cognitively taxing and can be especially difficult for older participants with neurological deficits. The heterogeneous nature of the lesions that patients present with introduces additional variability beyond pre-existing individual differences in cortical structure, which may further dilute subtle effects.

While these complications highlight various ways in which the data could be reconciled with IIT – or could turn out to be equivocal due to various sources of error – these experiments have the potential to significantly challenge the theory in two ways. First, and most obviously, both experiments could deliver compelling evidence that spatial experience is not warped in proximity to the blindspot or scotoma, despite the absence of particular neural architectures in corresponding cortical regions (consistent with the predictions of NREP and AI-C). Second, an inconsistent pattern of distortion might be observed for different kinds of task – e.g., a distance estimation task might reveal evidence of a contraction of phenomenal space, whereas an apparent motion task might suggest a dilation of phenomenal space (or vice-versa). Such mixed results would be difficult to explain solely in terms of structural connectivity (or predictive filling-in mechanisms). If this were to transpire, additional work would be required to explain how the hypothesised impact of deviant cortical structure interacts with other factors to produce task-specific effects.

## 3.4 Hypothesis #3: On the contribution of active sampling to changes in conscious visual experience

The final hypothesis we consider switches focus from empirical predictions motivated by IIT and NREP to the role of active sensory sampling as envisaged by AI-C. As discussed in Section 2.1.3, AI-C implies that some kind of (active) inference is required for a change in conscious experience to manifest – where action (i.e., the consequence of an inferred policy) here may be realised overtly (e.g., via motor behaviour) or covertly (e.g., via the reallocation of attention). This is in contrast to both IIT and NREP, neither of which accord any special role to policy selection in their accounts of consciousness. Although evidence in favour of the hypothesis that active sampling mediates changes in conscious content would not directly threaten the core claims of IIT or NREP, it would challenge their proponents to offer a satisfying explanation as to why such sampling should be expected to systematically influence the way perceptual contents manifest in consciousness.

As with the hypotheses discussed in the previous two sections, the putative role of active inference in conscious processing is difficult to test directly. Since active sampling is understood as a pervasive feature of living organisms under the Active Inference framework, it should not be possible to suppress all forms of such activity without rendering the organism unconscious (and perhaps even dead). As such, the hypothesis that active inference is necessary for a change in conscious content cannot simply be examined via the thoroughgoing suppression of all forms of active inference. Nonetheless, it should still be possible to shed light on this question by controlling the forms of action in which the human participant can engage.

The INTREPID Consortium aims to test the role of active sampling in mediating changes in consciousness via a modified version of the motion-induced blindness paradigm (Bonneh et al., 2001). In this task, participants report the subjective disappearance of a target stimulus presented in the visual periphery, as well as the subjective reappearance of the target after saccading towards it. The amount of time it takes for the individual to report subjective reappearance of the target stimulus during this active sampling condition will be compared to the time taken during a passive 'replay' condition in which the target stimulus (and the moving grid required to induce its disappearance) is moved to the centre of the visual field while the participant maintains stable fixation (rather than moving their eyes towards the peripheral location). In this way, the effect of active sampling (via saccade) on conscious content should be isolated while other factors (visual input, attentional deployment, task demands, etc.) remain constant (for further details, see Robinson et al., 2025).

AI-C predicts that active sampling of the target location should promote faster resolution of uncertainty about the latent causes of sensory states; hence, one should expect shorter subjective reappearance times to be reported as compared to the passive condition (which will still result in an eventual change in conscious content due to unresolved prediction errors generated by the new target position). Failure to confirm this prediction would present a significant challenge for AI-C, although it may not necessarily 'falsify' the theory outright. Auxiliary hypotheses that might explain the absence of the predicted effect could include (e.g.) technical difficulties (such as the inability to precisely match retinal input in the passive replay condition with that driven by eye movements performed during the active condition), measurement error or noise introduced by intervening processes between the onset of stimulus reappearance in conscious awareness and behavioural reports of this event, or some unanticipated confound that can be accommodated by the theoretical resources of the Active Inference framework (e.g., earlier subjective reappearance in the passive condition due to an increase in precision mediated by bottom-up attentional capture; later subjective reappearance in the active condition due to a decrease in precision mediated by top-down saccadic suppression). Of course, subsequent experiments could be devised with a view to disambiguating some of these candidate hypotheses, but they would themselves rely on further auxiliary hypotheses that could likewise be called upon to explain away evidence contrary to the predictions entailed by AI-C. This being said, successive failures to accrue evidence in favour of the theory would be indicative of a 'degenerating' research programme (i.e., one that requires multiple updates to account for existing data, yet which fail to deliver novel insights or predictions that are subsequently corroborated; Lakatos, 1976) – and reason to pursue alternative theories more vigorously (especially if they are accruing evidential support in the meantime).

As mentioned above, behavioural evidence favouring the facilitatory effect of action on changes in conscious contents would support AI-C, but would not necessarily constitute counter-evidence against IIT and NREP *per se*. Proponents of IIT might explain such an effect in terms of motor-induced changes in visual cortical state. Similarly, proponents of NREP might ascribe the effect to motor-driven changes in attention, or motor/reafference-related modulation of visual cortical dynamics (Oude Lohuis et al., 2024). However, while IIT and NREP are capable of accommodating (or explaining away) facilitatory effects of active sampling on conscious perception, they would do so by appealing to auxiliary hypotheses about the consequences of motor activity that are not straightforwardly derived from (or motivated by) their core claims. AI-C would by contrast afford a more compelling theoretical explanation of this phenomenon, insofar as it furnishes a principled, unifying account of the way changes in conscious experience are mediated by overt actions – and why such actions are accompanied by covert fluctuations in arousal and attention.

# 4 From empirical data to model comparison

Having implemented the experiments designed to test each of the hypotheses outlined in Section 3, the final step of our collaborative endeavour will be to assess the evidence in favour of each of the three theories of consciousness under examination. Although we are not in a position to comment on the precise ways in which each set of experimental findings will be analysed (since these decisions will be taken by the personnel in charge of each experiment), some general remarks on the potential advances, pitfalls, and controversies that may lie in wait can be made based on the state of the art of adversarial collaboration.

One notable empirical feature of adversarial collaboration is that it often fails to completely resolve scientific disputes amongst its protagonists (Mellers et al., 2001; Witkowski, 2020). Adversarial experiments typically generate mixed results that do not fully confirm the predictions of one theoretical position or another (as indeed was the case for the first ARC study; Cogitate Consortium et al., 2025). Even in cases where the data appear to decisively favour one set of predictions, proponents of the ‘disconfirmed’ hypothesis may ascribe unfavourable results to various auxiliary factors rather than concede the deficiency of their core theoretical claims (Cowan et al., 2020). As discussed in the previous section, this phenomenon is characteristic of scientific practice in general (i.e., it is not a distinctive feature of adversarial collaboration *per se*). Often, manuscripts reporting the results of an adversarial collaboration provide the opportunity for protagonists to outline their own interpretations of the results (Clark et al., 2022; Rakow, 2022). It is then up to the broader scientific community to evaluate these interpretations in light of the evidence, much as an impartial judge or jury must adjudicate the arguments presented by competing advocates in the adversarial legal system (Corcoran et al., 2023).

While the potential for divergent interpretations of results may not be unusual in adversarial collaboration, the multi-experiment structure of ARC projects such as INTREPID poses additional challenges for the adjudication of competing theories. The inclusion of multiple experiments raises the question of how the results of each experiment ought to be aggregated in view of the consideration that different experiments pose more or less severe tests of the theoretical predictions at stake. Since the ARC initiative stipulates that all experiments must be performed by two independent labs, there is also the additional question of how discrepancies in the results obtained by different labs performing ostensibly the same experiment should be handled (e.g., in the event that a predicted effect is reported by one lab but not the other, does this still constitute evidence in favour of the theory that predicted this effect, or should effects that 'fail to replicate' be demoted or disqualified?). Disagreements about the appropriate way to deal with such scenarios may engender further controversy and confusion about the implications of the data for competing theories.

These are challenging meta-scientific issues that transcend the scope of the INTREPID Consortium. However, since all ARC projects will ultimately need to contend with these issues (as will many other large-scale, multi-site collaborations), we briefly illustrate an analytic strategy that has recently been developed with a view to integrating the evidence accumulated across disparate datasets in the context of adversarial collaboration. The key idea of this approach – which forms part of a broader framework for Bayesian adversarial collaboration (Corcoran et al., 2023) – is to leverage standard techniques of variational Bayesian inference such that the hypotheses of competing theories can be formalised as generative models and fit to empirical data. The quality of these model fits – which is quantified as the (log) marginal likelihood or model evidence – can be obtained for each model and compared across theories (which may themselves be assigned different prior probabilities – although for the purposes of this Consortium the prior over each theory is assumed to be uniform). These quantities can be aggregated across datasets irrespective of whether they were generated by the same experiment, paradigm, or modality, meaning that the amount of evidence for (or against) each theory can be additively accrued as data are accumulated across different sites over the course of the project.

One advantage of this formal approach is that it encourages protagonists to specify their adversarial predictions as clearly and precisely as possible, thereby mitigating the risk of vague or ambiguous statements that can be retrospectively finessed to fit empirical results. This approach also facilitates the specification of competing predictions that are genuinely adversarial in the sense that they concern the same underlying data or putative effect, rather than targeting different phenomena that may not be mutually exclusive. This helps to ensure that the adversarial collaborative setting is being exploited to maximum effect (i.e., to drive an evidential wedge between theories by testing contrastive predictions about the same target phenomenon), rather than merely combining parallel investigations that do not directly engage with one another's hypotheses.

A further advantage of the integrative approach proposed by Corcoran and colleagues (2023) is its capacity to handle conflicting patterns of results (e.g., from different labs conducting the same protocol) in a principled manner. For instance, two implementations of the same experiment yielding equal amounts of evidence for and against the hypothesised effect would simply cancel each other out; epistemically speaking, one is no better off than before the experiments were run. However, if one lab produced strong evidence in favour of the effect, whereas another produced inconclusive evidence (perhaps due to noisier measurement), the data from the latter group would be down-weighted in accordance with its lower evidential value. Notably, such calculations would all be handled under the normative belief-updating scheme prescribed by Bayesian inference, thus obviating potentially controversial decisions pertaining to the amount of information conferred by disparate datasets (although decisions about which datasets ought to be included in the analysis would still require careful deliberation; cf. Negro et al., 2024).

To give a flavour of how the Bayesian model comparison scheme developed by Corcoran and colleagues (2023) might be used to evaluate the relative quantities of evidence accrued for IIT, NREP, and AI-C – over the course of the INTREPID Consortium – we provide a brief outline and some hypothetical examples designed to illustrate key features of the approach. Before doing so, it is worth highlighting two points: First, the adoption of a Bayesian scheme is motivated by practical rather than ideological considerations: Bayesian methods afford a rational way of accumulating evidence as new data are generated across experiments (and their replicates), where such evidence can support inferences about the existence or the absence of some phenomenon or effect (Spiegelhalter et al., 1994). In this way, Bayesian model comparison furnishes a probabilistic measure of the relative evidence accrued by competing theories; a measure that can be intuitively interpreted as the probability of alternative hypotheses or models, given the observations at hand. Although this approach does not set out to categorically endorse or refute specific theories, it affords a principled way of assessing the plausibility of competing theories based on the current state of the evidence.

The second point we underscore is how the strategy of fitting generative models for the purposes of theory comparison departs from more conventional applications in the service of parameter estimation. This may help to alleviate concerns about the sensitivity of Bayesian analysis to the specific form of the prior beliefs selected for model specification: that is, the concern that the selection of arbitrary priors may unduly bias the outcome of the analysis. In the present context, it is precisely the role of these priors to bias models towards one particular (set of) outcome(s); namely, that predicted by the specific theory that the model represents. In contrast to other settings – in which one seeks to specify models that are sufficiently responsive to novel data such that they converge on unbiased posterior estimates of underlying parameters – the intention here is to specify models that insist on how the data ought to appear under a given theory, and to evaluate the divergence between this conjectured reality and the way the data actually manifest. In this framework, stronger theoretical claims are encoded by less flexible (i.e., simpler) generative models, which are correspondingly less responsive to empirical observations (and thus run the risk of poorer model fits). Weaker claims, by contrast, are encoded by more expressive (i.e., complex) models that can accommodate a broader spectrum of potential observations. Intuitively, (the logarithm of) model evidence can always be expressed as accuracy minus complexity, where complexity is the divergence between posterior and prior estimates (i.e., after and before seeing the data). On this view, a weaker prior places less emphasis on model complexity and more emphasis on model accuracy and ensuing 'fit'.

With this analysis strategy in mind, each theory lead involved in the INTREPID project contributed their predictions for the effect of each critical manipulation on the key parameters of each experiment being conducted by the Consortium. These predictions took the form of a categorical 'checkbox' scheme, in which the predicted effect of an experimental manipulation or contrast could be positive, negative, or neutral (i.e., no change), relative to a control condition (see Tables 2-5). Theory leads were permitted to select more than one option, thereby encoding more complex predictions corresponding to directional or non-directional hypotheses, a hypothesis of no effect, or some combination of these options. Additionally, theory leads were invited to assign a categorical confidence level (low, medium, high) to each of their predictions. This information was solicited in recognition of the fact that some experiments test predictions that are more proximal to one theory than another; theorists may be less confident about the outcome of a manipulation involving factors that are more distal from or peripheral to the core claims of their theory. One can construe this confidence rating as hedging a kind of wager or 'Bayesian bet' (Corcoran et al., 2023; Tetlock & Mitchell, 2009b); namely, a way of weighting the evidence that will be accrued (or lost, in the case of an unpredicted outcome) from particular experiments. In this setting evidence pertaining to high-confidence predictions is up-weighted (i.e., a 'high-stakes' bet) and that pertaining to low-confidence predictions is down-weighted (i.e., a 'low-stakes' bet). Technically, the confidence plays the role of a precision or inverse temperature parameter that literally scales the (logarithm of the) evidence accrued by each theory.

A sample of the adversarial predictions elicited from each theory lead for each experiment is provided in Tables 2-5. Please note, due to space constraints we present a limited – but representative – subset of the predictions for each experiment; see the Consortium's preregistration (Olcese et al., 2024) and accompanying Study Protocols (Abbatecola et al., 2026; Haun et al., 2026; Robinson et al., 2025; Takahashi et al., 2025) for a more complete treatment. These tables show that predictions about the effect of each manipulation have been declared for all theories, ensuring that the results of these experiments will have evidential bearing on each theory. The tables also reveal cases in which predictions are (at least partially) shared across theories, where the confidence ascribed to the prediction may be the only point of distinction. While individual experiments may have limited power to disambiguate competing theories – due to overlapping expectations about the way the data will present – the aggregation of results over multiple experiments furnishes the opportunity for one theory to accumulate a considerable base of evidential support at the expense of its competitors.

<table>
<tr><td colspan="5"><b>Contrast: Optogenetic inactivation of inactive or minimally active neurons in left visual cortex (vs. no inactivation)</b></td></tr>
<tr><td></td><td>Biased towards right hemifield</td><td>No change in performance</td><td>Biased towards left hemifield</td><td>Confidence</td></tr>
<tr><td>IIT</td><td></td><td></td><td>X</td><td>High</td></tr>
<tr><td>NREP</td><td></td><td>X</td><td></td><td>High</td></tr>
<tr><td>AI-C</td><td></td><td>X</td><td></td><td>Medium</td></tr>
<tr><td colspan="5"><b>Contrast: Optogenetic inactivation of ‘normal background activity’ in left visual cortex (vs. no inactivation)</b></td></tr>
<tr><td></td><td>Biased towards right hemifield</td><td>No change in performance</td><td>Biased towards left hemifield</td><td>Confidence</td></tr>
<tr><td>IIT</td><td></td><td></td><td>X</td><td>High</td></tr>
<tr><td>NREP</td><td></td><td></td><td>X</td><td>Medium</td></tr>
<tr><td>AI-C</td><td>X</td><td></td><td>X</td><td>Low</td></tr>
</table>

**Table 2. Adversarial prediction tables for two experimental manipulations designed to test the effect of neural inactivation on visuospatial awareness.** IIT predicts both inactivations in the left visual cortex will bias responses on critical trials of a visual location task towards the intact (left) hemifield, consistent with the hypothesis that neural inactivation of already-inactive or minimally active neurons should alter the cause-effect structure in such a way as to impose a functional hemineglect. While NREP and AI-C agree amongst each other that the inactivation of minimally active neurons should not influence task performance – thus entailing the prediction of no change in behaviour relative to control trials in which the same task is performed in the absence of any optogenetic perturbation – NREP predicts an increase in response bias when ‘normal background activity’ is suppressed (see main text for definition of these activity levels). AI-C, by contrast, is less committal, offering a non-directional prediction about the consequence of this manipulation. Taken together, this set of predictions enables each of the three theories to accrue differing amounts of evidence according to the results of the experiment. See the preregistration (Olcese et al., 2024) and corresponding Study Protocol (Takahashi et al., 2025) for further details. AI-C: Active Inference theory of consciousness; IIT: Integrated Information Theory; NREP: Neurorepresentationalism.

| **Contrast: Presentation of punctate stimuli around blindspot (vs. away from blindspot)** | | | | |
|---|---|---|---|---|
| | Biased towards smaller distance | No change in performance | Biased towards larger distance | Confidence |
| IIT | X | | | Medium |
| NREP | X | X | | High |
| AI-C | | X | | Medium |
| | | | | |
| | Decreased sensitivity | No change in sensitivity | Increased sensitivity | Confidence |
| IIT | X | | | Medium |
| NREP | X | X | | Medium |
| AI-C | X | | | Medium |

**Table 3. Adversarial prediction table for an experimental manipulation designed to test the effect of the blindspot on psychometric functions derived from judgements of spatial extendedness.** IIT predicts stimuli spanning the observer's blindspot will appear closer together than identically spaced stimuli presented in an equivalent region of the visual field away from the observer's blindspot, resulting in a corresponding shift of the midpoint of the psychometric function. AI-C predicts no systematic difference in the subjective experience of the distance separating the two stimuli, irrespective of their location with respect to the blindspot. NREP predicts outcomes consistent with the individual predictions of both IIT and AI-C, but with higher confidence. All three theories predict diminished perceptual sensitivity when stimuli are presented around the blindspot, manifesting as a decrease in the slope of the psychometric function. NREP makes the additional prediction that sensitivity may remain unchanged relative to that observed for equivalent stimuli presented away from the blindspot. While no set of results would uniquely support the predictions of one theory while contradicting those of the other two, the differential weighting of hypotheses by confidence rating and complexity provide the opportunity for one theory to accrue more evidence than its competitors. See the preregistration (Olcese et al., 2024) and corresponding Study Protocol (Abbatecola et al., 2026) for further details. AI-C: Active Inference theory of consciousness; IIT: Integrated Information Theory; NREP: Neurorepresentationalism.

| **Contrast: Presentation of punctate stimuli around scotoma (vs. intact visual space)** | | | | |
|---|---|---|---|---|
| | Biased towards smaller distance | No difference in performance | Biased towards larger distance | Confidence |
| IIT | X | | | High |
| NREP | X | X | | High |
| AI-C | | X | | Medium |

**Table 4. Adversarial prediction table for an experiment designed to test the effect of scotoma on psychometric functions derived from judgements of spatial extendedness.** IIT predicts stimuli spanning or bordering cortical visual field defects (scotoma) will appear closer together to patients than compared to age-matched controls with no visual deficit. AI-C predicts no systematic difference in the subjective experience of the distance between stimuli, irrespective of whether the observer has a visual field defect or not. NREP predicts either no difference in task performance between patient and control groups (similar to AI-C), or that stimuli spanning scotoma will be experienced as closer together than they appear to observers without any visual deficit (similar to IIT). A result in favour of IIT or AI-C would also render evidence favouring NREP; however, NREP would accrue less evidence on account of its greater flexibility (i.e., staking two predictions instead of one). See the INTREPID Consortium's preregistration (Olcese et al., 2024) and corresponding Study Protocol (Haun et al., 2026) for further details. AI-C: Active Inference theory of consciousness; IIT: Integrated Information Theory; NREP: Neurorepresentationalism.

| **Contrast: Active (vs. passive) sampling of peripheral targets that have been suppressed from consciousness by motion-induced blindness** | | | | |
|---|---|---|---|---|
| | Faster time-to-reappearance | No change in performance | Slower time-to-reappearance | Confidence |
| IIT | X | X | | Low |
| NREP | X | X | | Medium |
| AI-C | X | | | High |

**Table 5. Adversarial prediction table for an experimental manipulation designed to test the effect of active sampling on the subjective reappearance of visual stimuli.** Both IIT and NREP predict that the time it will take for a visual stimulus to reappear in subjective awareness following a period of motion-induced blindness will not systematically differ across active (saccade) and passive (replay) conditions, or will be shorter in the active sampling condition. IIT stakes its predictions with lower confidence than NREP. AI-C commits to a directional hypothesis, predicting that active sampling will result in faster reappearance times. Evidence of slower reappearances in the active sampling condition would be costly for all three theories, with AI-C suffering the most severe loss of evidence due to its high-confidence prediction. See the INTREPID Consortium's preregistration (Olcese et al., 2024) and corresponding Study Protocol (Robinson et al., 2025) for further details. AI-C: Active Inference theory of consciousness; IIT: Integrated Information Theory; NREP: Neurorepresentationalism.

With a full set of prediction tables in hand, the next step is to formulate generative models characterising the data-generating process in each experiment: e.g., a binomial model capturing perceptual discrimination choices. Each model is then elaborated with theory-specific priors over key parameters: i.e., those pertaining to effects targeted in the prediction tables. These priors can be thought of as placing constraints on whether these parameters are free to vary (and if so, defining the domain over which variance is permitted). For instance, a parameter encoding the difference in rodent responses – during experimental trials involving optogenetic inactivation vs. control – could be fixed to zero, specifying the prediction that the suppression of visual cortical activity does not alter discrimination performance relative to that observed during the control condition. Conversely, this parameter could be allowed to vary over the positive domain, enabling the model to capture increases in response bias towards the inactivated side. These models are then inverted using standard variational methods (Fox & Roberts, 2012; Zeidman et al., 2023) to obtain estimates of their (log) model evidence (a.k.a., marginal likelihoods) via the optimisation of negative variational free energy (a.k.a., the evidence lower bound or ELBO; Winn & Bishop, 2005). This scores the difference between the accuracy of a model's predictions and its complexity. Finally, these quantities are scaled in accordance with the confidence rating ascribed to each prediction, such that the evidence gained for a predicted effect is boosted in the event that the prediction was wagered with greater precision. These quantities can simply be added together as more datasets become available, thus yielding updates of the evidential support accrued under each theory over time (for further details and worked examples, see Corcoran et al., 2023).

To illustrate these points more concretely, let us briefly consider a few hypothetical outcomes for the manipulations listed in Table 2. Let us first assume that optogenetic inactivation of inactive or minimally active neurons in mouse visual cortex produces no notable change in the rodents' pattern of responding. This outcome would constitute evidence against IIT and in favour of both predictive processing accounts; however, since NREP staked a strong Bayesian bet on the absence of an effect, it would benefit from greater evidence accumulation than AI-C, which tendered the same prediction with less confidence. If no notable change in choice behaviour were observed when normal background activity was suppressed, this would constitute evidence against all three theories; however, IIT would suffer the largest cost (loss of evidence), since it posited a prediction – for which there was no evidence – with the highest level of confidence. By contrast, AI-C would incur the smallest cost, since its unsupported predictions were staked with low confidence. Conversely, if the data were to reveal evidence of an effect (i.e., bias towards one or other hemifield), AI-C would only accrue a small amount of evidence in its favour, while IIT and NREP would gain (or lose) a larger amount depending on the directionality of the shift in response bias.

The Bayesian approach to evidence accumulation and model comparison outlined here is not expected to eliminate all sources of controversy from adversarial collaborative research; as indicated in Section 3, there are various reasons why the results bearing on a hypothesis might not turn out as anticipated – ranging from unforeseen experimental complications to under-appreciated causal factors – that may be elegantly integrated within an updated version of one's theory (or at least accommodated via the incorporation of auxiliary hypotheses). We see the strength of this analytic approach in its power to formalise predictions and confidences, quantify the evidence accumulated for competing theories, and integrate evidence over disparate paradigms and settings. Importantly, this analysis is envisaged as a foundation for substantive discussion – it does not supersede sensitive interpretation of the data-generating process or the need for carefully reasoned inferences to the best available explanation. We also acknowledge that this novel approach may introduce its own pitfalls, which may become apparent when deployed on real data. At the very least, it may be necessary to elaborate the scheme we propose here in a way that augments model comparison with procedures that can systematically account for the distance of a prediction from a theory's core (e.g., by weighting model evidences according to scores based on recently proposed metrics or criteria; see Chis-Ciure et al., 2024; Negro, 2024). It might also be useful to specify families of models permitted under competing theories, to account for subtle differences in predictions that issue from uncertainties in auxiliary hypotheses. These considerations aside, we are hopeful that this formal approach will complement and enrich the more informal commentaries typically provided in the discussion sections of adversarial collaborative research reports.

# 5 Concluding remarks

The Templeton World Charity Foundation's 'Accelerating Research on Consciousness' initiative marks an ambitious and unprecedented attempt to progress the neuroscience of consciousness through adversarial collaboration. Here, we have outlined the key tenets, similarities, and distinguishing features of three theories of consciousness – Integrated Information Theory, Neurorepresentationalism, and Active Inference – represented by one such adversarial collaboration – the INTREPID Consortium. We have furthermore identified key hypotheses that will be put to the test in a series of experiments designed to jointly elucidate which of these three theories provides the best account for a variety of empirical phenomena. Although we acknowledge that the adversarial model does not circumvent all of the challenges that attend the empirical validation of neuroscientific theories of consciousness, we suggest a Bayesian modelling framework may help to mitigate some of these difficulties by formally quantifying the amount of evidence accumulated for each theory across different experiments. We hope this strategy will be adopted and further developed in future work, with a view to facilitating the evidence-based adjudication of competing theories of consciousness in the community at large.

**Acknowledgements**

The authors wish to thank Clement Abbatecola, Jakob Hohwy, Belén Montabes de la Cruz, Umberto Olcese, and three anonymous reviewers for valuable discussion and feedback on previous versions of this manuscript.

**Funding**

This research was supported by a grant from the Templeton World Charity Foundation (TWCF#0646). The opinions expressed in this publication are those of the authors and do not necessarily reflect the views of the Templeton World Charity Foundation. AWC acknowledges the support of the Three Springs Foundation and the Lundbeck Foundation. KJF was supported by funding for the Wellcome Centre for Human Neuroimaging (Ref: 205103/Z/16/Z), a Canada-UK Artificial Intelligence Initiative (Ref: ES/T01279X/1) and the European Union's Horizon 2020 Framework Programme for Research and Innovation under the Specific Grant Agreement No. 945539 (Human Brain Project SGA3). CMAP was supported by the European Union's Horizon 2020 Framework Programme for Research and Innovation under the Specific Grant Agreement No. 945539 (Human Brain Project SGA3) and by NWO grant OCENW.M20.285.

**TWCF: INTREPID Consortium members**

Cyriel Pennartz, University of Amsterdam; Umberto Olcese, University of Amsterdam; Lars Muckli, University of Glasgow; Melanie Boly, University of Wisconsin-Madison; Jakob Hohwy, Monash University; Rafael Yuste, Columbia University; Patrick Cavanagh, York University Glendon College; Srimant Tripathy, University of Bradford; Eli Peli, Schepens Eye Research Institute of Harvard Medical School; Gyula Kovacs, Friedrich Schiller University Jena; Giulio Tononi, University of Wisconsin-Madison; Karl Friston, University College London; Kengo Takahashi, University of Amsterdam; Reinder Dorman, University of Amsterdam; Lucy Petro, University of Glasgow; Clement Abbatecola, University of Glasgow; Belen Montabes de la Cruz, University of Glasgow; Elena Monai, University of Wisconsin-Madison; Jonathan Robinson, Monash University; Christopher Whyte, Monash University; Samuel Pontes Quero, Columbia University; Marius 't Hart, York University Glendon College; Daejoon 'Alex' Hwang, Harvard University; Shiva Ram Male, Harvard University; Andras Sarkozy, Friedrich Schiller University Jena; Andrew Corcoran, Monash University; Kwangjun Lee, University of Amsterdam; Andrew Haun, University of Wisconsin-Madison; Lucia Melloni, Max Planck Institute; Anil Seth, University of Sussex; Nao Tsuchiya, Monash University; Nicholas Schiff, Weill Cornell Medical College; Wolf Singer, Max Planck Institute; Andy Clark, University of Sussex; Conrado Bosman, University of Amsterdam; Chris Klink, Netherland Institute for Neuroscience; Simon van Gaal, University of Amsterdam.